\documentclass[prl,a4paper,superscriptaddress,twocolumn,showpacs,amsmath,amssymb,floatfix]{revtex4}

\usepackage{graphicx}
\usepackage{amssymb}
\usepackage{float}

\usepackage{color}

\input{epsf}

\begin{document}

\title{Kolmogorov turbulence defeated by
Anderson localization \\
for a Bose-Einstein condensate in a Sinai-oscillator trap}

\author{Leonardo Ermann}
\affiliation{
   Departamento de F\'{\i}sica, Gerencia de Investigaci\'on y Aplicaciones,
 Comisi\'on Nacional de Energ\'{\i}a At\'omica.
 Av.~del Libertador 8250, 1429 Buenos Aires, Argentina}
\affiliation{CONICET, Godoy Cruz 2290 (C1425FQB) CABA, Argentina}
\author{Eduardo Vergini}
\affiliation{
   Departamento de F\'{\i}sica, Gerencia de Investigaci\'on y Aplicaciones,
 Comisi\'on Nacional de Energ\'{\i}a At\'omica.
 Av.~del Libertador 8250, 1429 Buenos Aires, Argentina}
\author{Dima L. Shepelyansky}
\affiliation{\mbox{Laboratoire de Physique Th\'eorique, IRSAMC, 
Universit\'e de Toulouse, CNRS, UPS, 31062 Toulouse, France}}

\date{March 21, 2017}

\begin{abstract}
We study the dynamics of a Bose-Einstein condensate
in a Sinai-oscillator trap under a monochromatic
driving force. Such a trap is formed 
by a harmonic potential and a repulsive disk 
located in the center vicinity corresponding to the
first experiments of condensate 
formation by Ketterle group in 1995.
We argue that the external driving
allows to model the regime of weak wave
turbulence with the Kolmogorov energy flow
from low to high energies.
We show that in a certain regime of
weak driving and weak nonlinearity
such a turbulent energy flow
is defeated by the Anderson localization
that leads to localization of energy on
low energy modes. A critical threshold
is determined
above which the turbulent flow to 
high energies becomes possible.
We argue that this phenomenon
can be studied with ultra cold atoms in
magneto-optical traps.
\end{abstract}

\pacs{05.45.Mt, 67.85.Hj,  47.27.-i, 72.15.Rn}
%

\maketitle

The Kolmogorov turbulence \cite{kolm41,obukhov}
is based on a concept of energy flow 
from large spacial scales, where an energy
is pumped by an external force,
to small scales where it is absorbed
by dissipation.
As a result a polynomial energy distribution
over wave modes is established
which has been obtained
first from scaling arguments 
for hydrodynamics turbulence \cite{kolm41,obukhov}.
Later the theory of weak turbulence, based on diagrammatic
techniques and the kinetic equation, 
demonstrated  the emergence of polynomial
distributions for various types of 
weakly interacting nonlinear waves
\cite{filonenko,zakharovbook,nazarenkobook}.
However, this theory is based on 
a fundamental hypothesis directly
stated in the seminal work of Zhakharov and Finonenko:
{\it ``In the theory of weak turbulence nonlinearity of waves is assumed to be
small; this enables us, using the hypothesis of the random nature of 
the phase of individual waves, to obtain the kinetic equation 
for the mean square of the wave amplitudes''}.
But in finite systems the dynamical equations
for waves do not involve Random Phase Approximation (RPA)
and the question of  RPA validity,
and hence of the whole concept
of energy flow from large to small scales,
remains open.

Indeed, it is known that in a random media
with a fixed potential landscape
the phenomenon of Anderson localization \cite{anderson1958}
breaks a diffusive spreading of probability
in space due to quantum interference effects,
even if the underline classical dynamics of particles
produces an unlimited spreading.
At present the Anderson localization
has been observed for a large variety
of linear waves in various physical systems \cite{akkermans}. 
A similar phenomenon appears also for
quantum systems in a periodically
driven {\it ac-}field with a quantum dynamical localization
in energy and number of absorbed photons 
\cite{chirikov1981,fishman1982,chirikov1988,prosen,dlsscholar}.
This dynamical localization in energy has been observed
in experiments with Rydberg atoms in a microwave field
\cite{koch,dlsscholar}
and cold atoms in driven optical lattices
\cite{raizen,garreau}.
Thus in the localized phase the periodic driving is not able to pump
energy to a system even if the classical dynamics is chaotic
with a diffusive spreading in energy.

Of course, the Anderson localization takes place for linear
waves. The question about its robustness in respect to a weak
nonlinearity attracted recently a significant interest of
nonlinear science community 
\cite{dlsdanse,pikodls,fishmandanse,ermannnjp,flach} with the
first experiments performed in  nonlinear media and optical lattices
\cite{segev,inguscio}.
These studies show that below
a certain threshold the Anderson localization
remains robust in respect to a weak nonlinearity
while above the threshold a subdiffusive
spearing over the whole system size
takes place. However, the studies are done for
conservative systems without external energy pumping.
The numerical simulations for a simple model of 
kicked nonlinear Schr\"odinger equation on a ring
gave indications that an energy flow to high
energies is stopped by the Anderson localization
for a weak nonlinearity \cite{dlskolm} but such a model
is rather far from real experimental 
possibilities with nonlinear media
or cold atoms.

In this Letter we consider a realistic system
of Bose-Einstein condensate (BEC) of cold atoms
captured in a Sinai-oscillator trap
under a monochromatic force.
In fact this system in three dimensions (3D)
had been used for a pioneering realization
of BEC reported in \cite{ketterle1995}
(see also \cite{ketterle2002,ketterle2002rmp}).
It represents a harmonic trap
with a repulsive potential in a vicinity
of the trap center 
created by a laser beam.
The repulsive potential can be well
approximated by a rigid disk
which creates scattering of atoms 
and instability of their classical dynamics.
In two dimensions (2D) with a harmonic
potential replaced by rigid
rectangular walls the systems 
represents the well-known Sinai billiard
where the mathematical results
guaranty that the whole system phase space is chaotic
with a positive Kolmogorov entropy \cite{sinai1970}.
Recently is was shown that the classical phase space  
remains practically fully chaotic
if the rigid walls are replaced by a harmonic potential
which is much more suitable for BEC experiments \cite{sinaiosl}.
The corresponding quantum system 
is characterized by the level spacing statistics
of random matrix theory \cite{wigner}
satisfying the Bohigas-Giannoni-Schmit conjecture \cite{bohigas}
and thus belonging to the systems of quantum chaos \cite{haake}.

The effects of nonlinearity for BEC evolution
in a Sinai-oscillator trap has been studied in \cite{sinaiosl}
in the frame of the Gross-Pitaevskii equation (GPE) \cite{becbook}.
It was shown \cite{sinaiosl}
that at weak nonlinearity the dynamics of linear modes
remains quasi-integrable while above a certain
threshold there is onset of dynamical thermalization 
leading to the usual Bose-Einstein distribution \cite{landau}
over energies of linear eigenmodes.
Even if being chaotic this system has energy conservation
and there is no energy flow to high energy modes.
However,  a monochromatic driving force
breaks the energy conservation leading 
for a classical dynamics to a diffusive
energy growth and probability transfer 
to high energy modes typical for the Kolmogorov turbulence.
Here we show that there is a regime where such
an energy transfer to waves with high wave vectors
is suppressed by the dynamical localization being similar
to the Anderson localization in disordered solids. 

We note that the Kolmogorov turbulence
for BEC in 2D rectangular and 3D cubic billiards
has been studied numerically in 
\cite{nazarenko2014,tsubota2015}.
However, the integrable shape of these
billiards does not allow to realize a generic
case of random matrix spectrum
of linear modes typical for our
billiard belonging
to the class of quantum chaos systems \cite{haake}.

For our model the classical dynamics and quantum evolution 
in absence of interactions
are described by the Hamiltonian
\begin{equation}
 \hat{H_0} =(\hat{p_x}^2+\hat{p_y}^2)/2m +
m(\omega_x^2 \hat{x}^2+\omega_y^2 \hat{y}^2)/2+
V_d(\hat{x},\hat{y})+f \hat{x} \sin{\omega t}  \;\; .
\label{eq1}
\end{equation}
Here the first two terms describe 2D oscillator with 
frequencies $\omega_x, \omega_y$,  the third term represents
the potential of rigid disk of radius  $r_d$ and the last term 
gives a driven monochromatic field of amplitude $f$.
Here we fixed the mass $m=1$, 
frequencies $\omega_x=1$, $\omega_y=\sqrt{2}$, $\omega=(1+\sqrt{5})$
and disk radius $r_d=1$. The disk center is placed at
 $(x_d,y_d)=(-1/2,-1/2)$ so that the disk bangs a hole in a center
vicinity as it was the case in the experiments \cite{ketterle1995}.
In the quantum case we have the usual commutator relations
$[\hat{p_x} \hat{x}] =  [\hat{p_y} \hat{y}] = - i \hbar$
with $\hbar=1$ for dimensional units.

 \begin{figure}[t]
\begin{center}
\includegraphics[width=0.42\textwidth]{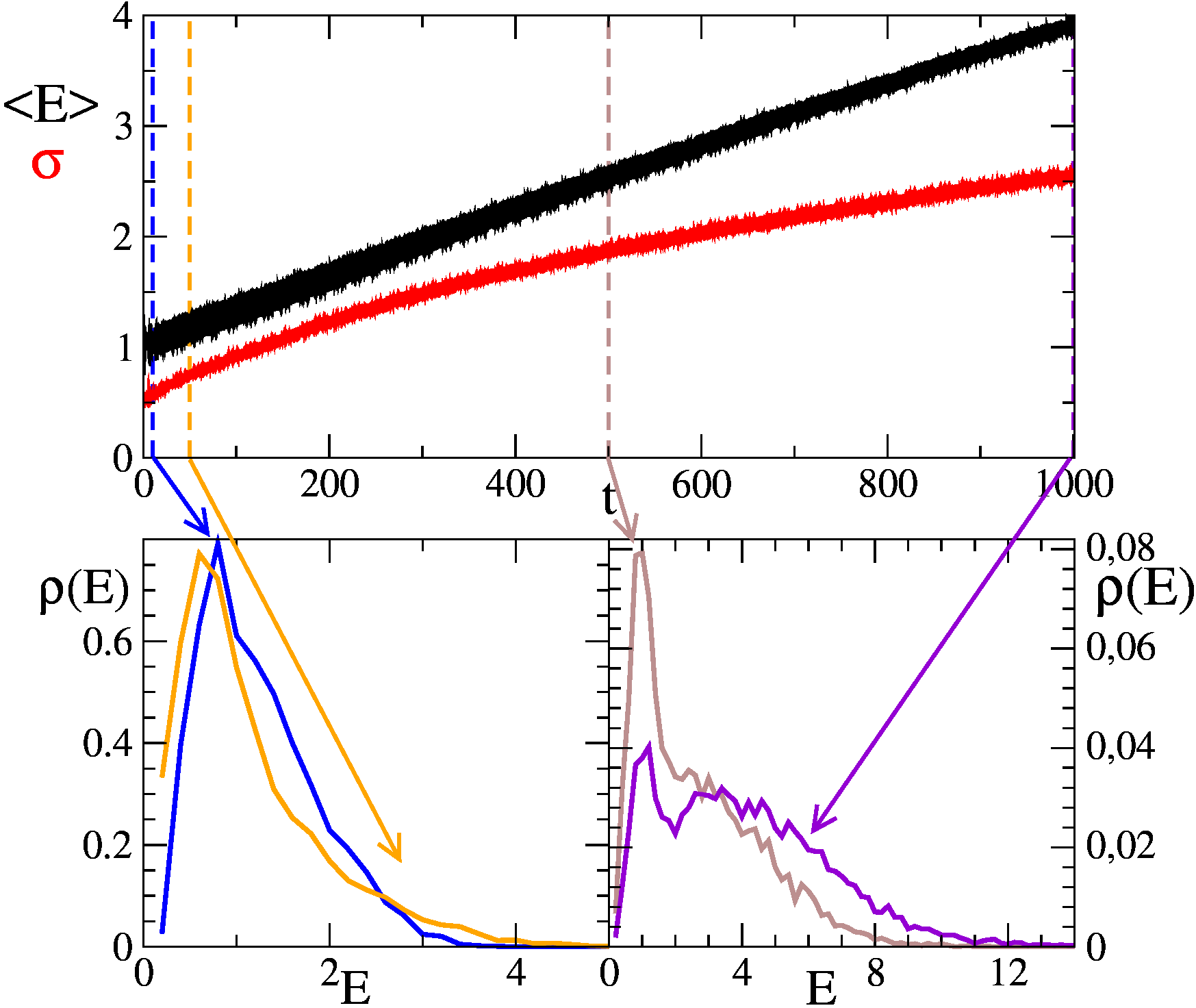}
\end{center}
\vglue -0.3cm
\caption{\label{fig1}
(Color online) Classical time evolution of 
average energy $<E>$  and its standard deviation $\sigma$ for $f=0.4$. 
The data are obtained from $10^4$ trajectories with random initial 
conditions at $<E>=1$ and $\sigma=0.5$. 
Top panel: $<E(t)>$ and $\sigma(t)$ are
shown by black and red/gray curves respectively. 
Bottom panels show probability distribution of trajectories $\rho(E,t)$ 
for  (a) $t=10,50$ (blue/black, orange/gray curves)  and 
(b) $t=500,1000$ 
(yellow/gray, violet/black curves). 
Vertical dashed lines in main panels mark snapshot times 
corresponding to 
bottom panels.
}
\end{figure}

The BEC evolution in the Sinai oscillator trap
is described by the GPE, which reads:
\begin{eqnarray}
\label{eq2}
 i\hbar{\partial\psi(x,y,t)\over\partial t} =  \hat{H_0}\psi(x,y,t) 
+   \beta \vert\psi(x,y,t)\vert^2\psi(x,y,t) \; ,
\end{eqnarray}
where $\beta$ describes nonlinear interactions for BEC.
Here we use the same Sinai oscillator parameters as in \cite{sinaiosl}
with normalization $\int |\psi|^2 dx dy =1$. 
The numerical integration of (\ref{eq2}) is done in the same 
way as in \cite{prosen,sinaiosl} with a Trotter time step
($\Delta t =0.005$) 
evolution for noninteracting part
of $\hat{H_0}$ followed by the nonlinear term contribution.

The results for energy $E$ growth 
with time for  classical dynamics (\ref{eq1})
are shown in Fig.~\ref{fig1}. 
The  energy $E$ and its dispersion
$\sigma$ are steadily growing with time.
We expect that at large times
the energy increases diffusively
with a rate $(\Delta E)^2/t = 
D \approx C f^2 {\omega_x}^2 r_d \sqrt{E}/\omega^2$
assuming that $\omega_x \sim \omega_y$ and $\omega > \omega_x$.
The data of Fig.~\ref{fig1} give us 
$C \approx 0.5$ at $t=10^3$. 

We note that the estimate for $D$ comes from the fact that an oscillating 
velocity component $v_{osc} = f \cos(\omega t)/\omega$
gives a velocity change at disk collision
(like with oscillating wall) $\Delta v_x = 2 v_{osc}$ and
an energy change $\Delta E \approx v_x \Delta v_x$
so that the diffusion is 
$D \sim (\Delta E)^2 /t_c$
where an average time between collisions
$t_c $ is defined from the ergodicity relation
$\Delta t_c /t_c  \sim {r_d}^2 {\omega_x}^2/E$
of ratio of  disk area and area of chaotic motion
at energy $E$, where $\Delta t_c \sim r_d/E^{1/2}$;
thus at large times $E \propto t^{2/3}$.
The fit for $E \sim \sigma \propto t^\alpha$ in Fig.~\ref{fig1}
gives $\alpha = 0.98 \pm 0.06$ (for $E$)
and $0.58 \pm 0.08$ (for $\sigma$)
being comparable to the theoretical value $\alpha=2/3$.
We  attributed a deviation from theory 
to not sufficiently large
amplitude of motion $\sqrt{2E}/\omega_x$
required for $t_c$ expression at reached energies.

We also introduce cells of finite energy size $\delta E$
and determine the probability distribution $\rho_k$  over $k$
energy cells counting a relative number of trajectories inside each cell.
The results of Fig.~\ref{fig1}  show that
the width of probability distribution $\rho(E)$ 
in  energy is growing in time
corresponding to increase of  $E$.

\begin{figure}[t]
\begin{center}
\includegraphics[width=0.48\textwidth]{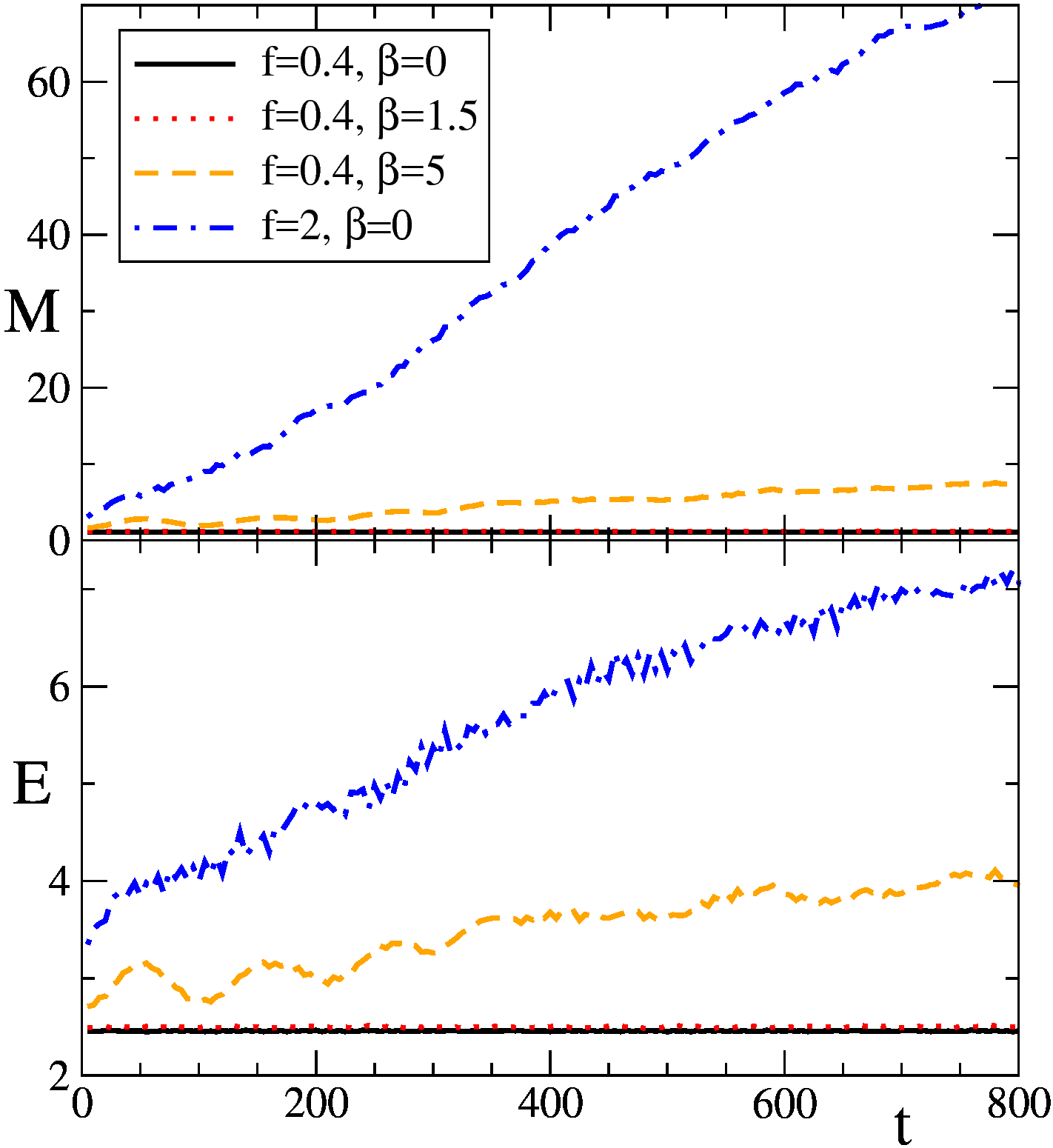}
\end{center}
\vglue -0.3cm
\caption{\label{fig2}
(Color online) Time evolution of $M$ (top panel) and energy $E$
(bottom panel) for GPE (\ref{eq2}) averaged over
time intervals  $\Delta t=1$.
The initial state is the ground state of (\ref{eq2}) 
at $\beta=0, f=0$
(see Fig.5a in \cite{sinaiosl}).
Both panels show the cases of $f=0.4, \beta=0$ (black solid lines),
$f=0.4, \beta=1.5$ (red/gray dotted lines),
$f=0.4, \beta=5$ (orange/gray dashed lines),
$f=2, \beta=0$ (blue/gray dot-dashed lines),
}
\end{figure}

\begin{figure}[t]
\begin{center}
\includegraphics[width=0.48\textwidth]{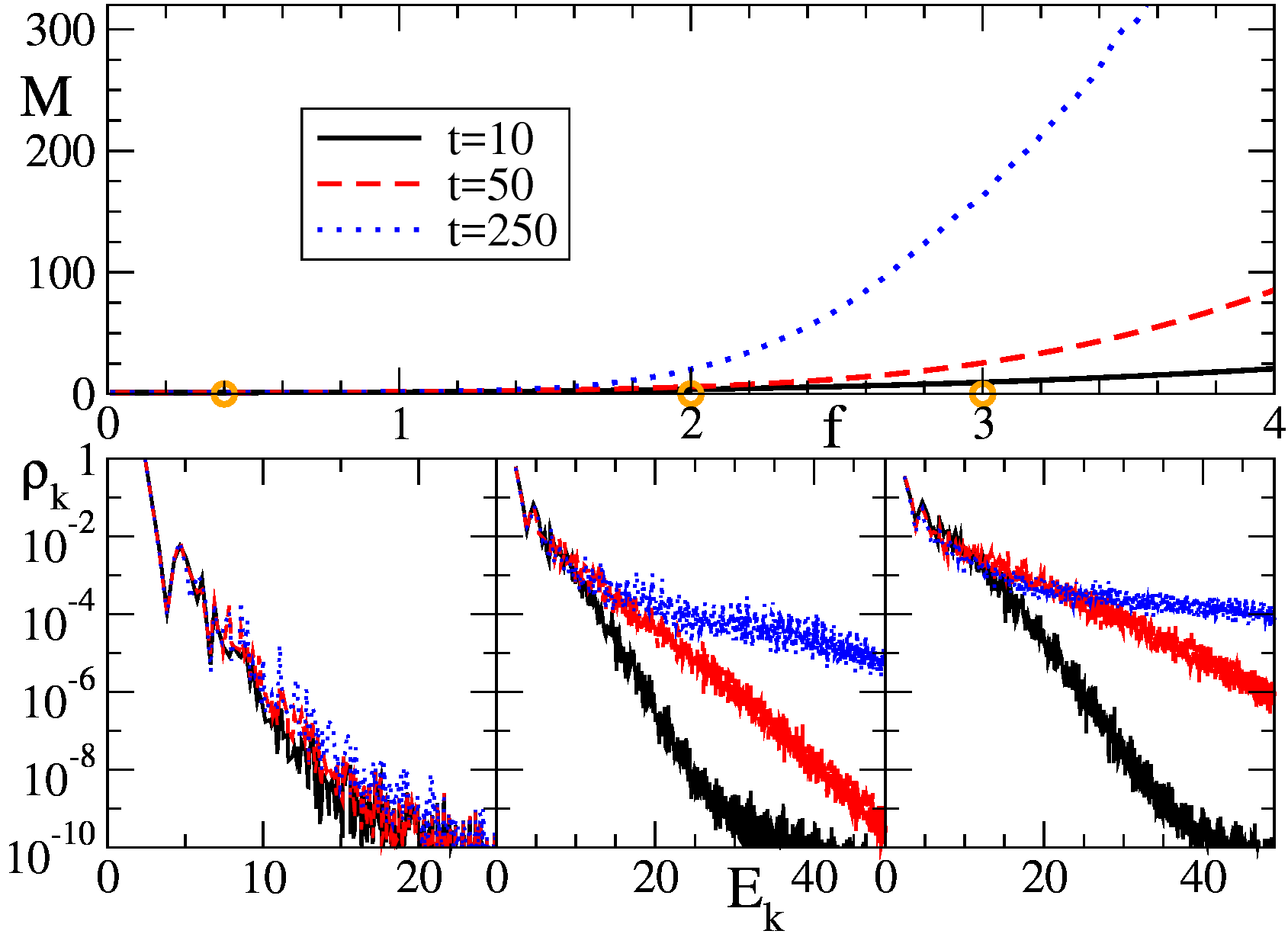}
\end{center}
\vglue -0.3cm
\caption{\label{fig3}
(Color online) 
Top panel shows $M$ as a function of driven force $f$ for linear case 
($\beta=0$).
Bottom panels show probability distribution $\rho_k$,
averaged over time interval $\Delta t=5$,
as a function of eigenenergies $E_k$ with $t=10$ in black solid lines,
$t=50$ in red/gray dashed lines, and $t=250$ in blue/gray dotted lines;
Left, center and right bottom panels show the cases of $f=0.4,2,3$ respectively 
(highlighted with orange/gray circles in top panel).
}
\end{figure}

The situation is drastically different in the quantum case at
$\beta=0$. Here, at small $f$, the dynamical localization leads to a complete
suppression of energy $E$ and average mode number $M= \sum_k k \rho_k$  growth
with  their restricted oscillations in time (see Fig.~\ref{fig2}).
The probability distribution $\rho_k$ 
over eigenstates  $\psi_k$ with eigenenergies $E_k$  of (\ref{eq1}) 
(for stationary case $f=0$) is shown in Fig.~\ref{fig3}.
For small $f< f_c$, on average there is a clear exponential decay
of probability  $\rho_k \propto \exp(- 2 E_k /\omega \ell_\phi)$
with a number of absorbed photons $N_\phi = E_k/\omega$
and a photonic localization length $\ell_\phi$.
Such a localization decay is similar to 
those discussed for atoms \cite{dlsscholar,koch}
and quantum dots \cite{prosen} in a microwave field.
However, above a certain $f_c$, e.g. at $f=2; 3$,
we obtain delocalized probabilities $\rho_k$ with a flat 
plateau distribution at high energies.

According to the theory of dynamical localization 
described in \cite{dlsscholar,prosen,deloc1d}
we have $\ell_\phi \approx 2\pi (D/\omega^2) \rho_c$
where $\rho_c=d k/dE_k$ is the density of $E_k$ states. 
According to \cite{sinaiosl}
we have $k \approx E^2/2\sqrt{2}$ and $\rho_c \approx E/\sqrt{2}$.
With the above expression for the classical diffusion in energy $D$
we obtain
$\ell_\phi  \approx  2 f^2 {\omega_x}^2 E^{3/2}/\omega^4$.
Similar to the quantum chaos model \cite{deloc1d}
we have $\ell_\phi$ significantly growing with 
the number of absorbed photons $N_\phi$ 
so that the delocalization of quantum chaos takes
place at $\ell_\phi > N_\phi$.
As in \cite{deloc1d} this leads to a delocalization
above a certain border $f>f_c$
with a flat probability distribution
on high energies as it is seen in Fig.~\ref{fig3}.
This gives the delocalization border for quantum states:
$f_c r_d/\hbar \omega_x \approx 0.7 (\omega/\omega_x)^{3/2} \approx 4$
for the initial ground state at 
$E \approx \hbar \omega_x =1$ and $\omega \approx 3.2$.
The data for $M$ in Fig.~\ref{fig3} give the critical value
$f_c \approx 1.5 $ being somewhat smaller than the value given by the
above estimate. We attribute this difference
to the fact that the above estimate for $D$, and hence for
$\ell_\phi$, is valid in the limit of large spacial oscillations being
larger than $r_d$.
The delocalization transition at $f>f_c$ is similar to the Anderson
transition, or metal-insulator transition, in
disordered systems \cite{anderson1958,akkermans}.

\begin{figure}[t]
\begin{center}
\includegraphics[width=0.48\textwidth]{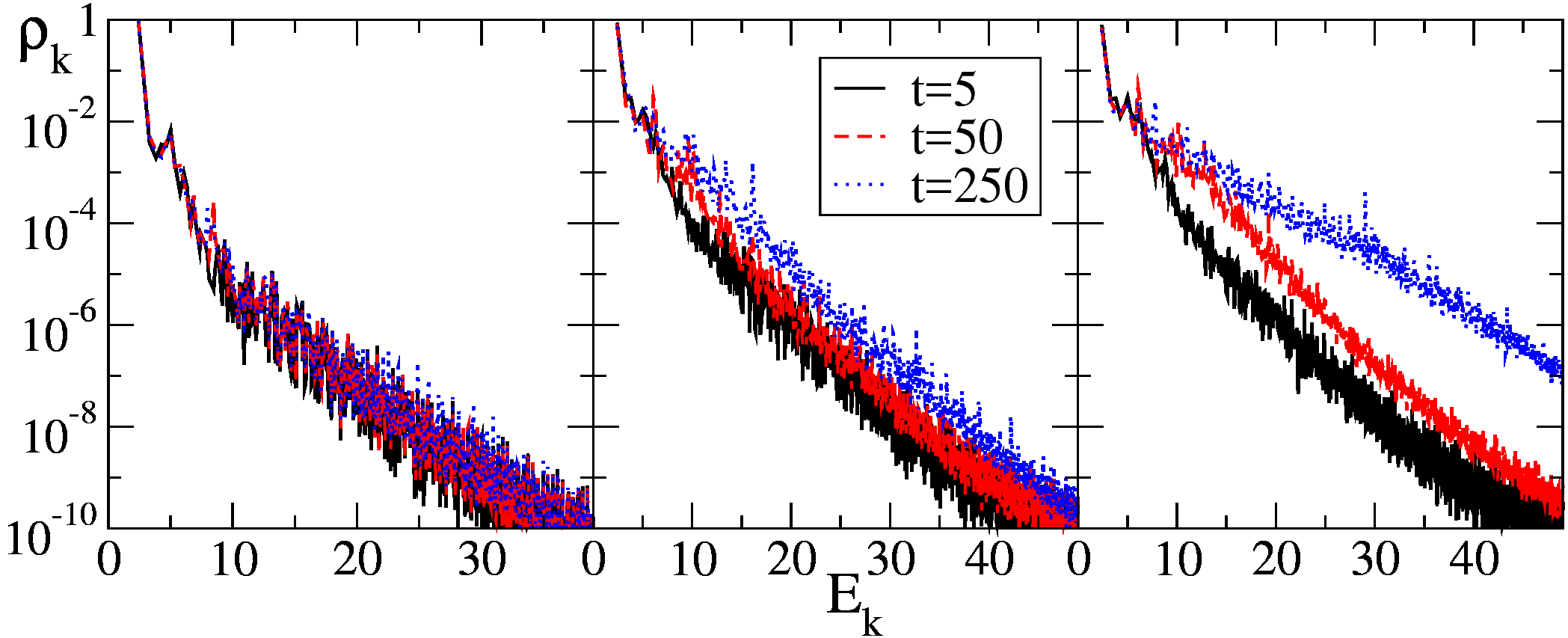}
\end{center}
\vglue -0.3cm
\caption{\label{fig4}
(Color online) Same as in bottom panels of Fig.~\ref{fig3}
for $f=0.3, \beta=1.5$ (left panel),
$f=0.5, \beta=5$ (center panel)
$f=1, \beta=5$ (right panel).
}
\end{figure}

The results for $\beta >0$ are presented in Figs.~\ref{fig2},~\ref{fig4}.
For $f=0.4$, when the 
steady-state probability is well localized 
at $\beta =0$, they clearly show that
at $\beta=1.5$  there is no growth 
of energy $E$ and mode number $M$.
Thus there is no energy flow to high energies
and the Anderson localization remains robust
for weak nonlineary perturbation.
This is also well confirmed by a stable in time
probability distribution over energies $E_k$
shown in Fig.~\ref{fig4} (left panel).
For larger nonlinearity $\beta=5$ and $f=0.4$
there appears a  growth of $M, E$ with time (Fig.~\ref{fig2}).
At larger $f=1$ and $\beta=5$
there is emergence of energy 
flow to high energies 
and increasing probability
$\rho_k$ at high energies $E_k$ 
(Fig.~\ref{fig4} right panel).

\begin{figure}[t]
\begin{center}
\includegraphics[width=0.48\textwidth]{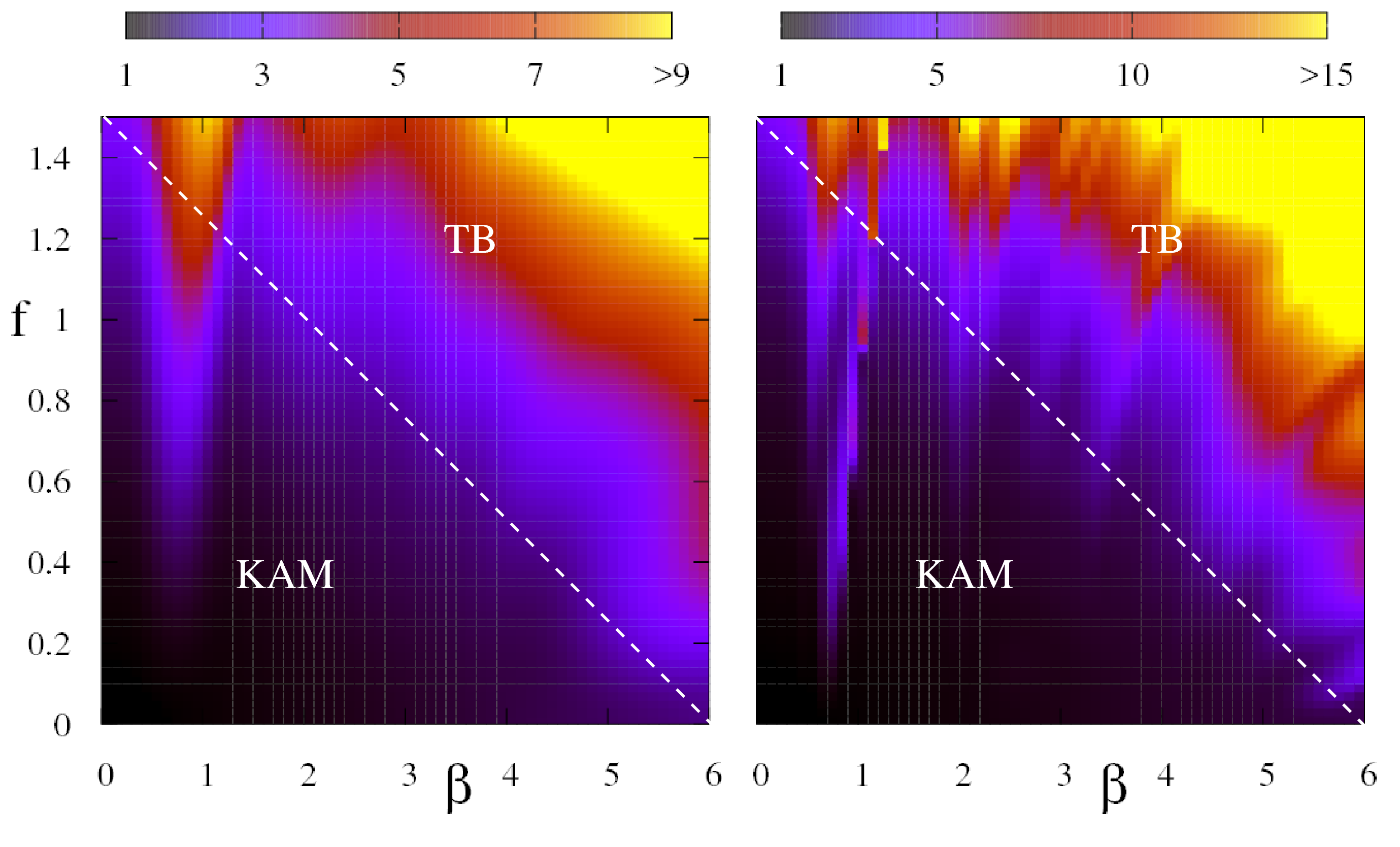}
\end{center}
\vglue -0.3cm
\caption{\label{fig5}
(Color online) Number of modes $M$ is 
shown by color/grayness in the plane of
parameters $f$ and $\beta$
(average is done in the time intervals
$100 \leq t \leq 150$ and
$250 \leq t \leq 300$
in left and right panel respectively.
The approximate separation 
of KAM or insulator phase (KAM)
and delocalized turbulent or  metallic phase
(TB) is shown by the white line 
(\ref{eq3}).
}
\end{figure}

The global dependence of average mode number $M$
on driving amplitude $f$ and nonlinearity $\beta$
is shown in Fig.~\ref{fig5}. We see that there is 
a stability region of small $f, \beta$ values where 
the values $M$  remain small even at
large times. This region corresponds
to the localized insulator phase (I), from the view point of
Anderson localization,
of quasi-integrable (or laminar)
phase from the view
point of nonlinear dinamics (or turbulence).
Outside of this region we 
have large values of number of populated states
$M$ so that this regime corresponds
to the delocalized metallic or turbulence phase
(M-TB).
According to the obtained results we conclude
that this quasi-stable  (or insulator) regime $(f<f_c,\beta<\beta_c)$
(see Fig.~\ref{fig5})
is approximately described by the relation
\begin{equation}
\label{eq3}
 f_c r_d / \hbar \omega_x \approx 1.5 (1 - \beta_c/(6 \hbar\omega_x {r_d}^2)) \; 
\end{equation}
assuming that $\omega_x \sim \omega_y \sim \omega$.
Inside the I-region
the turbulent Kolmogorov flow of energy to high modes 
is suppressed by the Anderson localization.
At small nonliniarity $\beta$ we expect a validity
of the Kolmogorov-Arnold-Moser theory (KAM)
\cite{chirikov,lichtenberg}  leading to
a quasi-integrable dynamics and trapping of energy on 
large length modes. At the same time we should
note that the mathematical prove of
KAM for nonlinear perturbation of pure-point
spectrum of Anderson localization
and the GPE (\ref{eq2}) still
remains an open challenge 
\cite{fishmandanse,wang,kuksin}.

Outside of the stability region (\ref{eq3})
a microwave driving transfers the energy flow
from low to high energy modes 
generating the Kolmogorov energy flow.
We expect that the energy dissipation
and high modes leads to the 
the Kolmogorov spectrum of energy distribution 
\cite{zakharovbook,nazarenkobook} over modes.
Our results show that the RPA is definitely not valid
and that, at small amplitudes of a monochromatic
driving and small nonlinearity,
the Kolmogorov turbulent flow to high modes is
defeated by the Anderson localization 
and the KAM integrability. The transition from KAM
phase to turbulence phase
corresponds to the insulator-metal transition
in disordered systems with the energy
axis corresponding to the spatial distance
respectively. 
The KAM or insulator phase corresponds
to a usual observation that a small wind
(small $f$ amplitude)
is not able to generate turbulent waves.

The experimental realization of our
system  with BEC in a magneto-optical trap
corresponds to the experimental conditions
described in \cite{ketterle1995}.
A monochromatic perturbation can be created
by oscillations of the center of harmonic potential
or effectively by  oscillations
of the disk position created by the laser beam.
We note that the experimental investigations 
of turbulent cascades in quantum gases  
become now possible \cite{becturbulence}
as well as a thermometry of energy distribution
in ultra cold atom ensembles \cite{grimm}.
Thus we hope that the interesting fundamental 
aspects of nonlinear dynamics and weak turbulence 
will be tested with cold atom experiments.

This work was supported in part by the Pogramme Investissements
d'Avenir ANR-11-IDEX-0002-02, 
reference ANR-10-LABX-0037-NEXT (project THETRACOM).


\end{document}